\documentclass[11pt]{article}
\usepackage{amsmath,amssymb,amsthm,amsxtra,overpic,bbm,bm,epsfig}
\def\thefootnote{\fnsymbol{footnote}}

\begin{document}

\vspace{0.2cm}

\begin{center}
{\Large\bf Properties of CP Violation in Neutrino-Antineutrino
Oscillations} \footnote{This paper is dedicated to Bruno Pontecorvo,
who has been considered the father of neutrino-antineutrino
oscillations, on the occasion of his 100th birthday this year (22
August 2013). It is also dedicated to the 40th birthday of my home
institute, the Institute of High Energy Physics, which was founded
on 1 February 1973.}
\end{center}

\vspace{0.1cm}

\begin{center}
{\bf Zhi-zhong Xing}
\footnote{E-mail: xingzz@ihep.ac.cn} \\
{Institute of High Energy Physics, Chinese Academy of
Sciences, P.O. Box 918, Beijing 100049, China \\
}
\end{center}

\vspace{1.5cm}

\begin{abstract}
If the massive neutrinos are the Majorana particles, how to pin down
the Majorana CP-violating phases will eventually become an
unavoidable question relevant to the future neutrino experiments. We
argue that a study of neutrino-antineutrino oscillations will
greatly help in this regard, although the issue remains purely
academic at present. In this work we first derive the probabilities
and CP-violating asymmetries of neutrino-antineutrino oscillations
in the three-flavor framework, and then illustrate their properties
in two special cases: the normal neutrino mass hierarchy with
$m^{}_1 =0$ and the inverted neutrino mass hierarchy with $m^{}_3
=0$. We demonstrate the significant contributions of the Majorana
phases to the CP-violating asymmetries, even in the absence of the
Dirac phase.
\end{abstract}

\begin{flushleft}
\hspace{0.8cm} PACS number(s): 14.60.Pq, 13.10.+q, 25.30.Pt \\
\hspace{0.8cm} Keywords: Majorana neutrino, neutrino-antineutrino
oscillation, CP violation
\end{flushleft}

\def\thefootnote{\arabic{footnote}}
\setcounter{footnote}{0}

\newpage

If the massive neutrinos are the Majorana particles, then a neutrino
flavor $\nu^{}_\alpha$ can in principle oscillate into an
antineutrino flavor $\overline{\nu}^{}_\beta$ (for $\alpha, \beta =
e, \mu, \tau$). The intriguing idea of neutrino-antineutrino
oscillations was first proposed by Pontecorvo in 1957
\cite{Pontecorvo}, but it has been regarded to be unrealistic
because such lepton-number-violating processes are formidably
suppressed by the factors $m^2_i/E^2$ with $m^{}_i \lesssim 1$ eV
(for $i=1,2,3$) being the neutrino masses and $E$ being the neutrino
beam energy \cite{Valle1981}. Taking the reactor antineutrino
experiment for example, one has $E \sim {\cal O}(1)$ MeV and thus
$m^2_i/E^2 \lesssim 10^{-12}$, implying that the probability of
$\overline{\nu}^{}_e \to \nu^{}_e$ oscillations is too small to be
observable. That is why only the phenomena of neutrino-neutrino and
antineutrino-antineutrino oscillations, which are
lepton-number-conserving and do not involve the helicity suppression
factors $m^2_i/E^2$, have so far been observed in solar,
atmospheric, reactor and accelerator experiments \cite{PDG}. If the
Majorana nature of the massive neutrinos is identified someday, will
it be likely to detect neutrino-antineutrino oscillations in a
realistic experiment?

The answer to this question seems to be quite pessimistic today, but
it might not be really hopeless in the future. The history of
neutrino physics is full of surprises in making the impossible
possible. Let us mention a naive idea. To enhance the helicity
suppression factors $m^2_i/E^2$, one may consider to invent some new
techniques and produce a sufficiently low energy neutrino (or
antineutrino) beam. For instance, the possibility of producing a
M$\rm\ddot{o}$ssbauer electron antineutrino beam with $E = 18.6$ keV
\cite{M1}
\footnote{The M$\rm\ddot{o}$ssbauer electron antineutrinos are the
18.6 keV $\overline{\nu}^{}_e$ events emitted from the bound-state
beta decay of $^3{\rm H}$ to $^3{\rm He}$ \cite{Bahcall}, and they
can be resonantly captured in the reverse bound-state process in
which $^3{\rm He}$ is converted into $^3{\rm H}$.}
and using it to do an $\overline{\nu}^{}_e \to \overline{\nu}^{}_e$
oscillation experiment has been discussed \cite{M2}. If the
$\overline{\nu}^{}_e \to \nu^{}_e$ oscillation is taken into account
in this case, the helicity suppression can be improved by a factor
of ${\cal O}(10^4)$ as compared with the case of the aforementioned
reactor antineutrinos.

It is theoretically interesting to study the properties of
neutrino-antineutrino oscillations even in a Gedanken experiment,
because they may help understand some salient properties of the
Majorana neutrinos. This kind of study has been done in the
literature \cite{Valle1981,Kayser}, but in most cases only two
species of neutrinos and antineutrinos were taken into account.

In the present work we shall first derive the probabilities of
neutrino-antineutrino oscillations within the standard three-flavor
framework, and then discuss the generic properties of CP violation
in them. To illustrate, we shall focus on the CP-violating effects
in neutrino-antineutrino oscillations by considering two special
cases of the neutrino mass spectrum: (a) the normal hierarchy with
$m^{}_1 = 0$; and (b) the inverted hierarchy with $m^{}_3 = 0$. We
demonstrate the importance of the Majorana phases in generating the
CP-violating asymmetries, even when the Dirac phase is absent. Our
analytical results can easily be generalized to accommodate the
light or heavy sterile Majorana neutrinos and antineutrinos.

\vspace{0.4cm}

Let us begin with the standard form of leptonic weak charged-current
interactions:
\begin{eqnarray}
{\cal L}^{}_{\rm cc} \ = \ -\frac{g}{\sqrt 2} \left[\overline{\left(
\begin{matrix} e & \mu & \tau \end{matrix} \right)^{}_{\rm L}} ~
\gamma^\mu \ U \left(\begin{matrix} \nu^{}_1 \cr \nu^{}_2 \cr
\nu^{}_3 \cr
\end{matrix} \right)^{}_{\rm L} W^-_\mu ~ + ~
\overline{\left( \begin{matrix} \nu^{}_1 & \nu^{}_2 & \nu^{}_3
\end{matrix} \right)^{}_{\rm L}} ~ \gamma^\mu \ U^\dagger
\left(\begin{matrix} e \cr \mu \cr \tau \cr \end{matrix}
\right)^{}_{\rm L} W^+_\mu \right] \; ,
\end{eqnarray}
in which $U$ is the $3\times 3$ Pontecorvo-Maki-Nakagawa-Sakata
(PMNS) flavor mixing matrix \cite{MNSP}. Now we consider
$\nu^{}_\alpha \to \nu^{}_\beta$ and $\nu^{}_\alpha \to
\overline{\nu}^{}_\beta$ oscillations (for $\alpha, \beta = e, \mu,
\tau$), whose typical Feynman diagrams are illustrated in Figure 1.
It is clear that the $\nu^{}_\alpha \to \nu^{}_\beta$ oscillations
are lepton-number-conserving and can take place no matter whether
the massive neutrinos are the Dirac or Majorana particles. In
contrast, the $\nu^{}_\alpha \to \overline{\nu}^{}_\beta$
oscillations are lepton-number-violating and cannot take place
unless the massive neutrinos are the Majorana particles. Focusing on
the oscillation $\nu^{}_\alpha \to \overline{\nu}^{}_\beta$ and its
CP-conjugate process $\overline{\nu}^{}_\alpha \to \nu^{}_\beta$,
one may write out their amplitudes as follows
\cite{Valle1981,Kayser}
\footnote{Here we do not consider the details on the production of
$\nu^{}_\alpha$ (or $\overline{\nu}^{}_\alpha$) and the detection of
$\overline{\nu}^{}_\beta$ (or $\nu^{}_\beta$), and thus it is
possible to factorize the amplitudes of $\nu^{}_\alpha \to
\overline{\nu}^{}_\beta$ and $\overline{\nu}^{}_\alpha \to
\nu^{}_\beta$ as in Eq. (2) \cite{Khalil}.}
\begin{eqnarray}
A(\nu^{}_\alpha \to \overline{\nu}^{}_\beta) \hspace{-0.15cm} & = &
\hspace{-0.15cm} \sum_i \left[ U^*_{\alpha i} U^*_{\beta i}
\frac{m^{}_i}{E} \exp\left(-{\rm
i}\frac{m^2_i}{2E} L\right) \right] K \; , \nonumber \\
A(\overline{\nu}^{}_\alpha \to \nu^{}_\beta) \hspace{-0.15cm} & = &
\hspace{-0.15cm} \sum_i \left[ U^{}_{\alpha i} U^{}_{\beta i}
\frac{m^{}_i}{E} \exp\left(-{\rm i}\frac{m^2_i}{2E} L\right) \right]
\overline{K} \; ,
\end{eqnarray}
where $m^{}_i$ is the mass of the neutrino mass eigenstate
$\nu^{}_i$, $E$ denotes the neutrino (or antineutrino) beam energy,
$L$ is the baseline length, $K$ and $\overline{K}$ stand for the
kinematical factors which are independent of the index $i$ (and
satisfy $|K| = |\overline{K}|$). The helicity suppression in the
transition between $\nu^{}_i$ and $\overline{\nu}^{}_i$ is described
by $m^{}_i/E$, which is absent for normal neutrino-neutrino or
antineutrino-antineutrino oscillations.
\begin{figure}[t]
\centering \vspace{-0.15cm}
\includegraphics[width=.63\textwidth]{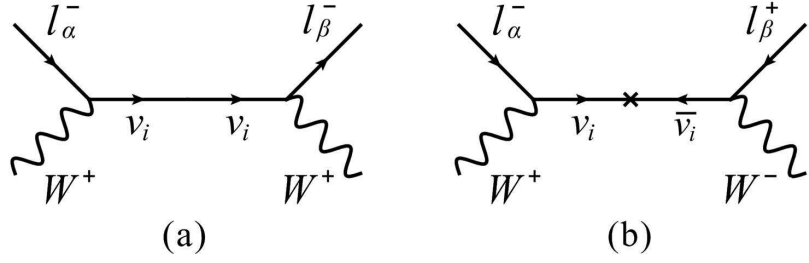}
\caption{Feynman diagrams for (a) neutrino-neutrino and (b)
neutrino-antineutrino oscillations, where ``$\times$" stands for the
chirality flip in the neutrino propagator which is proportional to
the mass $m^{}_i$ of the Majorana neutrino $\nu^{}_i =
\overline{\nu}^{}_i$. The initial ($\nu^{}_\alpha$) and final
($\nu^{}_\beta$ or $\overline{\nu}^{}_\beta$) neutrino flavor
eigenstates are produced and detected via the weak charged-current
interactions, respectively.}
\end{figure}

Eq. (2) allows us to calculate the probabilities of
neutrino-antineutrino oscillations $P(\nu^{}_\alpha \to
\overline{\nu}^{}_\beta) \equiv |A(\nu^{}_\alpha \to
\overline{\nu}^{}_\beta)|^2$ and $P(\overline{\nu}^{}_\alpha \to
\nu^{}_\beta) \equiv |A(\overline{\nu}^{}_\alpha \to
\nu^{}_\beta)|^2$. After a straightforward exercise, we arrive at
\begin{eqnarray}
P(\nu^{}_\alpha \to \overline{\nu}^{}_\beta) \hspace{-0.15cm} & = &
\hspace{-0.15cm} \frac{|K|^2}{E^2} \left[ \left| \langle m
\rangle^{}_{\alpha \beta} \right|^2 - 4 \sum_{i<j} m^{}_i m^{}_j
{\rm Re}\left( U^{}_{\alpha i} U^{}_{\beta i} U^*_{\alpha j}
U^*_{\beta j} \right) \sin^2 \frac{\Delta m^2_{ji} L}{4E} \right .
\nonumber \\
&& \left . \hspace{2.5cm} + 2 \sum_{i<j} m^{}_i m^{}_j {\rm Im}
\left ( U^{}_{\alpha i} U^{}_{\beta i} U^*_{\alpha j} U^*_{\beta j}
\right) \sin \frac{\Delta m^2_{ji} L}{2E} \right] \; ,
\nonumber \\
P(\overline{\nu}^{}_\alpha \to \nu^{}_\beta) \hspace{-0.15cm} & = &
\hspace{-0.15cm} \frac{|\overline{K}|^2}{E^2} \left[ \left| \langle
m \rangle^{}_{\alpha \beta} \right|^2 - 4 \sum_{i<j} m^{}_i m^{}_j
{\rm Re}\left( U^{}_{\alpha i} U^{}_{\beta i} U^*_{\alpha j}
U^*_{\beta j} \right) \sin^2 \frac{\Delta m^2_{ji} L}{4E} \right .
\nonumber \\
&& \left . \hspace{2.5cm} - 2 \sum_{i<j} m^{}_i m^{}_j {\rm Im}
\left ( U^{}_{\alpha i} U^{}_{\beta i} U^*_{\alpha j} U^*_{\beta j}
\right) \sin \frac{\Delta m^2_{ji} L}{2E} \right] \; ,
\end{eqnarray}
in which $\Delta m^2_{ji} \equiv m^2_j - m^2_i$, and the effective
mass term $\langle m \rangle^{}_{\alpha \beta}$ is defined as
\begin{eqnarray}
\langle m \rangle^{}_{\alpha \beta} \ \equiv \ \sum_i m^{}_i
U^{}_{\alpha i} U^{}_{\beta i} \ \equiv \ M^{}_{\alpha\beta} \; ,
\end{eqnarray}
which is simply the $(\alpha, \beta)$ element of the Majorana
neutrino mass matrix $M = U \widehat{M} U^T$ with $\widehat{M}
\equiv {\rm Diag}\{m^{}_1, m^{}_2, m^{}_3\}$ in the flavor basis
where the charged-lepton mass matrix is diagonal \cite{Xing}. The
CPT invariance assures that $P(\nu^{}_\alpha \to
\overline{\nu}^{}_\beta) = P(\nu^{}_\beta \to
\overline{\nu}^{}_\alpha)$ and $P(\overline{\nu}^{}_\alpha \to
\nu^{}_\beta) = P(\overline{\nu}^{}_\beta \to \nu^{}_\alpha)$ hold.
The CP-violating asymmetry between $\nu^{}_\alpha \to
\overline{\nu}^{}_\beta$ and $\overline{\nu}^{}_\alpha \to
\nu^{}_\beta$ oscillations turns out to be
\begin{eqnarray}
{\cal A}^{}_{\alpha\beta} \ \equiv \ \frac{P(\nu^{}_\alpha \to
\overline{\nu}^{}_\beta) - P(\overline{\nu}^{}_\alpha \to
\nu^{}_\beta)}{P(\nu^{}_\alpha \to \overline{\nu}^{}_\beta) +
P(\overline{\nu}^{}_\alpha \to \nu^{}_\beta)} \ = \
\frac{\displaystyle 2 \sum_{i<j} m^{}_i m^{}_j {\rm Im} \left (
U^{}_{\alpha i} U^{}_{\beta i} U^*_{\alpha j} U^*_{\beta j} \right)
\sin \frac{\Delta m^2_{ji} L}{2E}}{\displaystyle \left| \langle m
\rangle^{}_{\alpha \beta} \right|^2 - 4 \sum_{i<j} m^{}_i m^{}_j
{\rm Re}\left( U^{}_{\alpha i} U^{}_{\beta i} U^*_{\alpha j}
U^*_{\beta j} \right) \sin^2 \frac{\Delta m^2_{ji} L}{4E}} \; ,
\end{eqnarray}
which is no more suppressed by $m^2_i/E^2$. Of course,
${\cal A}^{}_{\alpha\beta} = {\cal A}^{}_{\beta\alpha}$ holds too.
Hence only six of the nine CP-violating asymmetries are independent.
Eqs. (3) and (5) allow us to look at the salient features of
neutrino-antineutrino oscillations and CP violation in them. Some
discussions are in order.

(a) {\it The zero-distance effect}. Taking $L = 0$, one obtains
\begin{eqnarray}
P(\nu^{}_\alpha \to \overline{\nu}^{}_\beta) \ = \
P(\overline{\nu}^{}_\alpha \to \nu^{}_\beta) \ = \ \frac{|K|^2}{E^2}
\left| \langle m \rangle^{}_{\alpha \beta} \right|^2 \; ,
\end{eqnarray}
which is CP-conserving (i.e., ${\cal A}^{}_{\alpha\beta} =0$ at
$L=0$). Given $\alpha = \beta = e$, for example, the above
probabilities are actually determined by the effective mass term
$|\langle m\rangle^{}_{ee}|$ of the neutrinoless double beta decay.
A measurement of the latter will therefore provide a meaningful
constraint on the oscillation between electron neutrinos and
electron antineutrinos. Of course, the zero-distance effect in Eq.
(6) is extremely suppressed due to $E \gg |\langle
m\rangle^{}_{\alpha\beta}|$ in practice. Note that $P(\nu^{}_\alpha
\to \nu^{}_\beta) = P(\overline{\nu}^{}_\alpha \to
\overline{\nu}^{}_\beta) = \delta^{}_{\alpha\beta}$ holds at $L =0$
in normal neutrino-neutrino or antineutrino-antineutrino
oscillations, provided $U$ is unitary.

(b) {\it CP violation in $\nu^{}_\alpha \to
\overline{\nu}^{}_\alpha$ oscillations}. We find that Eq. (3) will
not be much simplified even if $\alpha = \beta$ is taken, and the
CP-violating term will not disappear in this case. The point is
simply that the $\nu^{}_\alpha \to \overline{\nu}^{}_\alpha$
oscillation is actually a kind of ``appearance" process, different
from the normal $\nu^{}_\alpha \to \nu^{}_\alpha$ and
$\overline{\nu}^{}_\alpha \to \overline{\nu}^{}_\alpha$ oscillations
which belong to the ``disappearance" processes. In this
flavor-unchanging case,
\begin{eqnarray}
{\cal A}^{}_{\alpha\alpha} \ = \ \frac{\displaystyle 2 \sum_{i<j}
m^{}_i m^{}_j {\rm Im} \left ( U^{2}_{\alpha i} U^{* 2}_{\alpha j}
\right) \sin \frac{\Delta m^2_{ji} L}{2E}}{\displaystyle \left|
\langle m \rangle^{}_{\alpha \alpha} \right|^2 - 4 \sum_{i<j} m^{}_i
m^{}_j {\rm Re}\left( U^{2}_{\alpha i} U^{*2}_{\alpha j} \right)
\sin^2 \frac{\Delta m^2_{ji} L}{4E}} \; .
\end{eqnarray}
Of course, ${\cal A}^{}_{\alpha\alpha}$ (or more generally, ${\cal
A}^{}_{\alpha\beta}$) may vanish on the ``finely tuned" points with
$\Delta m^2_{ji} L/(2E) = \pi, 2\pi, 3\pi$, and so on. But such
special points can only be chosen, in principle, for a monochromatic
neutrino or antineutrino beam \cite{Kayser}.

(c) {\it The Majorana CP-violating phases}. As shown in Eq. (3) or
Eq. (5), the effects of CP violation in neutrino-antineutrino
oscillations are measured by ${\rm Im} (U^{}_{\alpha i} U^{}_{\beta
i} U^*_{\alpha j} U^*_{\beta j})$, which would vanish if the PMNS
matrix $U$ were real. The combination $U^{}_{\alpha i} U^{}_{\beta
i} U^*_{\alpha j} U^*_{\beta j}$ is invariant under a redefinition
of the phases of three charge-lepton fields, but it is sensitive to
the rephasing of the neutrino fields
\footnote{In comparison, the strength of CP violation in normal
neutrino-neutrino or antineutrino-antineutrino oscillations is
determined by ${\rm Im} (U^{}_{\alpha i} U^{}_{\beta j} U^*_{\alpha
j} U^*_{\beta i})$ \cite{J}, which is absolutely
rephasing-invariant. In other words, it is impossible to probe the
Majorana nature of the massive neutrinos (or antineutrinos) through
the $\nu^{}_\alpha \to \nu^{}_\beta$ (or $\overline{\nu}^{}_\alpha
\to \overline{\nu}^{}_\beta$) oscillations.}.
Hence the Majorana CP-violating phases of $U$ must play an important
role in neutrino-antineutrino oscillations via ${\rm Im}
(U^{}_{\alpha i} U^{}_{\beta i} U^*_{\alpha j} U^*_{\beta j})$, even
if $\alpha = \beta$ is taken. This observation motivates us to ask
such a meaningful question: what can we do about the Majorana
CP-violating phases after the Majorana nature of the massive
neutrinos is identified via a measurement of the neutrinoless double
beta decay \cite{Barger} and the Dirac CP-violating phase is determined
through a delicate long-baseline experiment of neutrino oscillations
in the foreseeable future? The experiment of neutrino-antineutrino
oscillations is apparently a possible way towards pinning down or
constraining the Majorana CP-violating phases, although it is
considerably challenging. Is there a better way out?

\vspace{0.4cm}

To see the properties of CP violation (or equivalently, the roles of
the Majorana phases) in neutrino-antineutrino oscillations in a
simpler and clearer way, let us take two phenomenologically allowed
limits of the neutrino mass spectrum for illustration.

(1) {\it A special normal mass hierarchy with $m^{}_1 = 0$}. In this
case the $3\times 3$ PMNS matrix $U$ can be parametrized in terms of
three mixing angles ($\theta^{}_{12}, \theta^{}_{13},
\theta^{}_{23}$) and two CP-violating phases ($\delta, \sigma$)
\cite{Mei}:
\begin{eqnarray}
U \ = \ \left( \begin{matrix} c^{}_{12} c^{}_{13} & s^{}_{12}
c^{}_{13} & s^{}_{13} e^{-{\rm i} \delta} \cr -s^{}_{12} c^{}_{23} -
c^{}_{12} s^{}_{13} s^{}_{23} e^{{\rm i} \delta} & c^{}_{12}
c^{}_{23} - s^{}_{12} s^{}_{13} s^{}_{23} e^{{\rm i} \delta} &
c^{}_{13} s^{}_{23} \cr s^{}_{12} s^{}_{23} - c^{}_{12} s^{}_{13}
c^{}_{23} e^{{\rm i} \delta} & -c^{}_{12} s^{}_{23} - s^{}_{12}
s^{}_{13} c^{}_{23} e^{{\rm i} \delta} & c^{}_{13} c^{}_{23} \cr
\end{matrix} \right) \left(
\begin{matrix} 1 & 0 & 0 \cr 0 & e^{{\rm i} \sigma} & 0 \cr 0 & 0 & 1
\cr \end{matrix} \right) \; ,
\end{eqnarray}
where $c^{}_{ij} \equiv \cos\theta^{}_{ij}$ and $s^{}_{ij} \equiv
\sin\theta^{}_{ij}$ (for $ij = 12, 13, 23$). A global analysis of
the available neutrino oscillation data \cite{FIT} points to
$\theta^{}_{12} \simeq 33.4^\circ$, $\theta^{}_{13} \simeq
8.66^\circ$ and $\theta^{}_{23} \simeq 40.0^\circ$, but $\delta$ is
essentially unrestricted. In addition, $m^{}_2 = \sqrt{\Delta
m^2_{21}} \simeq 8.66 \times 10^{-3}$ eV and $m^{}_3 = \sqrt{\Delta
m^2_{31}} \simeq 4.97 \times 10^{-2}$ eV are obtained by using the
typical inputs $\Delta m^2_{21} \simeq 7.50 \times 10^{-5} ~{\rm
eV}^2$ and $\Delta m^2_{31} \simeq 2.47 \times 10^{-3} ~{\rm eV}^2$
\cite{FIT}. Both $\delta$ and $\sigma$ enter the CP-violating
asymmetry ${\cal A}^{}_{\alpha\beta}$, which is now simplified to
\begin{eqnarray}
{\cal A}^{}_{\alpha\beta} \hspace{-0.15cm} & = & \hspace{-0.15cm}
\frac{\displaystyle 2 m^{}_2 m^{}_3 {\rm Im} \left ( U^{}_{\alpha 2}
U^{}_{\beta 2} U^*_{\alpha 3} U^*_{\beta 3} \right) \sin
\frac{\Delta m^2_{32} L}{2E}}{\displaystyle \left| m^{}_2
U^{}_{\alpha 2} U^{}_{\beta 2} + m^{}_3 U^{}_{\alpha 3} U^{}_{\beta
3} \right|^2 - 4 m^{}_2 m^{}_3 {\rm Re}\left( U^{}_{\alpha 2}
U^{}_{\beta 2} U^*_{\alpha 3} U^*_{\beta 3} \right) \sin^2
\frac{\Delta m^2_{32} L}{4E}}
\nonumber \\
\hspace{-0.15cm} & = & \hspace{-0.15cm} \frac{\displaystyle 2 {\rm
Im} \left ( U^{}_{\alpha 2} U^{}_{\beta 2} U^*_{\alpha 3} U^*_{\beta
3} \right) \sin \frac{\Delta m^2_{32} L}{2E}}{\displaystyle \left|
\sqrt{\frac{m^{}_2}{m^{}_3}} U^{}_{\alpha 2} U^{}_{\beta 2} +
\sqrt{\frac{m^{}_3}{m^{}_2}} U^{}_{\alpha 3} U^{}_{\beta 3}
\right|^2 - 4 {\rm Re}\left( U^{}_{\alpha 2} U^{}_{\beta 2}
U^*_{\alpha 3} U^*_{\beta 3} \right) \sin^2 \frac{\Delta m^2_{32}
L}{4E}} \; .
\end{eqnarray}
We see that the ratio $\sqrt{m^{}_2/m^{}_3} \simeq 0.42$ or its
reciprocal may more or less affect the magnitude of ${\cal
A}^{}_{\alpha\beta}$. The latter also depends on $\Delta m^2_{32}$
via its oscillating term.

(2) {\it A special inverted mass hierarchy with $m^{}_3 = 0$}. In
this case the $3\times 3$ PMNS matrix $U$ can also be parametrized
as in Eq. (8) with a single Majorana CP-violating phase $\sigma$,
and the present global fit yields $\theta^{}_{12} \simeq
33.4^\circ$, $\theta^{}_{13} \simeq 8.66^\circ$ and $\theta^{}_{23}
\simeq 50.4^\circ$ \cite{FIT}. Furthermore, we obtain $m^{}_1 =
\sqrt{-\Delta m^2_{21} - \Delta m^2_{32}} \simeq 4.85 \times
10^{-2}$ eV and $m^{}_2 = \sqrt{-\Delta m^2_{32}} \simeq 4.93 \times
10^{-2}$ eV by using the typical inputs $\Delta m^2_{21} \simeq 7.50
\times 10^{-5} ~{\rm eV}^2$ and $\Delta m^2_{32} \simeq -2.43 \times
10^{-3} ~{\rm eV}^2$ \cite{FIT}. The CP-violating asymmetry ${\cal
A}^{}_{\alpha\beta}$ turns out to be
\begin{eqnarray}
{\cal A}^{}_{\alpha\beta} \hspace{-0.15cm} & = & \hspace{-0.15cm}
\frac{\displaystyle 2 m^{}_1 m^{}_2 {\rm Im} \left ( U^{}_{\alpha 1}
U^{}_{\beta 1} U^*_{\alpha 2} U^*_{\beta 2} \right) \sin
\frac{\Delta m^2_{21} L}{2E}}{\displaystyle \left| m^{}_1
U^{}_{\alpha 1} U^{}_{\beta 1} + m^{}_2 U^{}_{\alpha 2} U^{}_{\beta
2} \right|^2 - 4 m^{}_1 m^{}_2 {\rm Re}\left( U^{}_{\alpha 1}
U^{}_{\beta 1} U^*_{\alpha 2} U^*_{\beta 2} \right) \sin^2
\frac{\Delta m^2_{21} L}{4E}}
\nonumber \\
\hspace{-0.15cm} & \simeq & \hspace{-0.15cm} \frac{\displaystyle 2
{\rm Im} \left ( U^{}_{\alpha 1} U^{}_{\beta 1} U^*_{\alpha 2}
U^*_{\beta 2} \right) \sin \frac{\Delta m^2_{21}
L}{2E}}{\displaystyle \left| U^{}_{\alpha 1} U^{}_{\beta 1} +
U^{}_{\alpha 2} U^{}_{\beta 2} \right|^2 - 4 {\rm Re}\left(
U^{}_{\alpha 1} U^{}_{\beta 1} U^*_{\alpha 2} U^*_{\beta 2} \right)
\sin^2 \frac{\Delta m^2_{21} L}{4E}} \; ,
\end{eqnarray}
where $m^{}_1 \simeq m^{}_2$ has been adopted in obtaining the
approximate result. One can see that the magnitude of ${\cal
A}^{}_{\alpha\beta}$ is essentially independent of the absolute
neutrino masses $m^{}_1$ and $m^{}_2$ in this special case, although
it relies on $\Delta m^2_{21}$ via the oscillating term.
\begin{table}[t]
\caption{The CP-violating asymmetry of neutrino-antineutrino
oscillations in two special cases: (1) the normal neutrino mass
hierarchy with $m^{}_1 =0$ and $\Delta m^2_{32} L/(2E) = \pi/2$,
together with the typical inputs $\theta^{}_{12} \simeq 33.4^\circ$,
$\theta^{}_{13} \simeq 8.66^\circ$, $\theta^{}_{23} \simeq
40.0^\circ$, $\Delta m^2_{21} \simeq 7.50 \times 10^{-5} ~{\rm
eV}^2$ and $\Delta m^2_{31} \simeq 2.47 \times 10^{-3} ~{\rm eV}^2$;
(2) the inverted neutrino mass hierarchy with $m^{}_3 = 0$ and
$\Delta m^2_{21} L/(2E) = \pi/2$, together with the typical inputs
$\theta^{}_{12} \simeq 33.4^\circ$, $\theta^{}_{13} \simeq
8.66^\circ$, $\theta^{}_{23} \simeq 50.4^\circ$, $\Delta m^2_{21}
\simeq 7.50 \times 10^{-5} ~{\rm eV}^2$ and $\Delta m^2_{32} \simeq
-2.43 \times 10^{-3} ~{\rm eV}^2$. The typical values of the
CP-violating phases $\delta$ and $\sigma$ are taken below.}
\begin{center}
\begin{tabular}{llll}
\hline\hline Normal hierarchy \hspace{1.6cm} & $\delta = 0$ and
$\sigma =\pi/4$ \hspace{1.8cm} & $\delta = \pi/2$ and $\sigma =\pi/4$ \\
\hline
${\cal A}^{}_{ee}$ & $+0.74$ & $-0.74$ \\
${\cal A}^{}_{e\mu}$ & $+0.87$ & $+0.075$ \\
${\cal A}^{}_{e\tau}$ & $-0.80$ & $+0.088$ \\
${\cal A}^{}_{\mu\mu}$ & $+0.29$ & $+0.34$ \\
${\cal A}^{}_{\mu\tau}$ & $-0.25$ & $-0.25$ \\
${\cal A}^{}_{\tau\tau}$ & $+0.22$ & $+0.17$ \\
\hline Inverted hierarchy \hspace{1.6cm} & $\delta = 0$ and
$\sigma =\pi/4$ \hspace{1.8cm} & $\delta = \pi/2$ and $\sigma =\pi/4$ \\
\hline
${\cal A}^{}_{ee}$ & $-0.73$ & $-0.73$ \\
${\cal A}^{}_{e\mu}$ & $+0.91$ & $+0.92$ \\
${\cal A}^{}_{e\tau}$ & $+0.96$ & $+0.96$ \\
${\cal A}^{}_{\mu\mu}$ & $-1.00$ & $-0.54$ \\
${\cal A}^{}_{\mu\tau}$ & $-0.80$ & $-0.75$ \\
${\cal A}^{}_{\tau\tau}$ & $-0.46$ & $-0.64$ \\
\hline\hline
\end{tabular}
\end{center}
\end{table}

To illustrate the magnitude of ${\cal A}^{}_{\alpha\beta}$, one may
simplify its expression by taking $\Delta m^2_{32} L/(2E) = \pi/2$
in Eq. (9) or taking $\Delta m^2_{21} L/(2E) = \pi/2$ in Eq. (10).
In either case, it is now possible to get a ball-park feeling about
the size of ${\cal A}^{}_{\alpha\beta}$ if the values of the
CP-violating phases $\delta$ and $\sigma$ are input. For simplicity,
we fix $\sigma = \pi/4$ and take $\delta = 0$ or $\pi/2$. The
numerical results of ${\cal A}^{}_{\alpha\beta}$ are then listed in
Table 1
\footnote{For the inverted hierarchy with
$m^{}_3 = 0$, $\delta =0$ and $\sigma =\pi/4$, the result
${\cal A}^{}_{\mu\mu} \simeq -1.00$ in Table 1 is a
consequence of ${\rm Re}(U^2_{\mu 1}U^{*2}_{\mu 2}) = 0$, ${\rm
Im}(U^2_{\mu 1}U^{*2}_{\mu 2}) \simeq -|U^{}_{\mu 1}|^4$ and
$|U^2_{\mu 1} + U^2_{\mu 2}|^2 \simeq |U^{}_{\mu 1}|^4$ because of
the special values of the input parameters.}.
We see that there can be quite sizable CP-violating effects in
neutrino-antineutrino oscillations, and they may simply arise from
the Majorana CP-violating phase(s) even if the Dirac CP-violating
phase $\delta$ is switched off (or the flavor mixing angle
$\theta^{}_{13}$ is switched off).

In addition to the above two special cases, the three neutrinos may
also have a nearly degenerate mass spectrum \cite{FX}. Namely,
$m^{}_1 \approx m^{}_2 \approx m^{}_3$, but the exact equality is
forbidden because it is in conflict with the neutrino oscillation data.
In this interesting case, $m^{}_i \approx m^{}_j$ can be factored out
on the right-hand side of Eq. (3) and thus the CP-violating asymmetry
${\cal A}^{}_{\alpha\beta}$ in Eq. (5) is simplified to
\begin{eqnarray}
{\cal A}^{}_{\alpha\beta} \ \approx \
\frac{\displaystyle 2 \sum_{i<j} {\rm Im} \left (
U^{}_{\alpha i} U^{}_{\beta i} U^*_{\alpha j} U^*_{\beta j} \right)
\sin \frac{\Delta m^2_{ji} L}{2E}}{\displaystyle
\left| \sum_i U^{}_{\alpha i} U^{}_{\beta i} \right|^2 - 4 \sum_{i<j}
{\rm Re}\left( U^{}_{\alpha i} U^{}_{\beta i} U^*_{\alpha j}
U^*_{\beta j} \right) \sin^2 \frac{\Delta m^2_{ji} L}{4E}} \; ,
\end{eqnarray}
which is independent of the absolute neutrino masses. The values of
${\cal A}^{}_{\alpha\beta}$ may be sensitive to the sign of $\Delta
m^2_{31}$ (or $\Delta m^2_{32}$) through the sum of three
oscillating terms in the numerator of ${\cal A}^{}_{\alpha\beta}$.

Note that a complete or partial degeneracy of three neutrino masses
is sometimes taken to reveal the distinct properties of flavor
mixing and CP violation for the Majorana neutrinos \cite{Branco}. A
systematic analysis \cite{Mei2} shows that the PMNS matrix $U$ can
in general be simplified if both the neutrino mass degeneracy and
the Majorana phase degeneracy, which are both conceptually
interesting, are assumed. Given $m^{}_1 = m^{}_2 = m^{}_3$, for
example, Eq. (3) is simplified to
\begin{eqnarray}
P(\nu^{}_\alpha \to \overline{\nu}^{}_\beta) \ = \
P(\overline{\nu}^{}_\alpha \to \nu^{}_\beta) \ = \ \frac{|K|^2}{E^2}
m^2_1 \left| \sum_i U^{}_{\alpha i} U^{}_{\beta i} \right|^2 \; .
\end{eqnarray}
This result is very similar to the zero-distance effect given in Eq.
(6). Of course, ${\cal A}^{}_{\alpha \beta} =0$ holds in this
special case, although there are still nontrivial CP violating
phases in $U$. If the Majorana phases of three neutrinos were
exactly degenerate (i.e., $\phi^{}_1 = \phi^{}_2 = \phi^{}_3$ with
$\phi^{}_i$ being associated with the neutrino mass eigenstate
$\nu^{}_i$), we would be able to rotate away all of them from the
PMNS matrix $U$. In this case, the CP-violating asymmetry ${\cal
A}^{}_{\alpha\beta}$ is only dependent on the Dirac phase $\delta$.
This point can be clearly seen from the combination $U^{}_{\alpha i}
U^{}_{\beta i} U^*_{\alpha j} U^*_{\beta j}$ that appears in Eqs.
(3) and (5) \cite{next}.

\vspace{0.4cm}

In summary, we have derived the probabilities and CP-violating
asymmetries of neutrino-antineutrino oscillations in the standard
three-flavor framework
\footnote{Although it is impossible for the Dirac neutrinos to have
neutrino-antineutrino oscillations, it is possible for them to
oscillate between their left-handed and right-handed states in a
magnetic field and in the presence of matter effects (see Ref.
\cite{Giunti} for a review of such spin-flavor precession
processes).}.
We have particularly illustrated the CP-violating effects in
neutrino-antineutrino oscillations by considering two
phenomenologically allowed limits of the neutrino mass spectrum: (a)
the normal hierarchy with $m^{}_1 = 0$; and (b) the inverted
hierarchy with $m^{}_3 = 0$. The importance of the Majorana phases
in generating the CP-violating asymmetries, even when the Dirac
phase is absent, has been demonstrated. Our analytical results can
easily be generalized to accommodate the light or heavy sterile
Majorana neutrinos and antineutrinos.

We reiterate that this work is motivated by a meaningful question
that we have asked ourselves: what can we proceed to do to pin down the
full picture of flavor mixing and CP violation if the massive
neutrinos are identified to be the Majorana particles via a
convincing measurement of the neutrinoless double beta decay in the
future? By then we might be able to find a better way out
\footnote{Some new techniques for producing and measuring neutrinos
and antineutrinos, such as the one using atoms or molecules \cite{Petcov},
will probably be developed in the future.},
or just pay more attention to the feasibility of detecting
neutrino-antineutrino oscillations and CP violation in them.

\vspace{0.4cm}

I would like to thank Y.F. Li and S. Zhou for their interesting comments
and Y.L. Zhou for his help in plotting the original version of Figure 1.
This work is supported in part by the National Natural Science
Foundation of China under Grant No. 11135009.


\begin{thebibliography}{99}
\bibitem{next} We shall calculate the explicit expressions of
${\rm Re}(U^{}_{\alpha i} U^{}_{\beta i} U^*_{\alpha j} U^*_{\beta
j})$ and ${\rm Im}(U^{}_{\alpha i} U^{}_{\beta i} U^*_{\alpha j}
U^*_{\beta j})$ based on an explicit parametrization of $U$
elsewhere, so as to see the explicit dependence of ${\cal
A}^{}_{\alpha \beta}$ on $\delta$ in this special case (with
vanishing Majorana CP-violating phases).
\end{thebibliography}
\end{document}